# A numerical approach for hybrid reliability analysis of structures under mixed uncertainties using the uncertainty theory

Lei Zhang[a,b]

[a] Beihang University, Beijing 100083, China

[b] Leibniz Universität Hannover, Hannover 30167, Germany

**Abstract**: This paper presents a novel numerical method for the hybrid reliability analysis by using the uncertainty theory. Aleatory uncertainty and epistemic uncertainty are considered simultaneously in this method. Epistemic uncertainty is characterized by the uncertainty theory, and the effect of epistemic uncertainty is quantified by the sub-additive uncertain measure. Then, under the framework of the chance theory which can be interpreted as the combination of the probability theory and the uncertainty theory, a general uncertainty quantification model is established to deal with the hybrid reliability analysis problem, then the corresponding reliability metric is defined. After that, to improve the feasibility of the proposed model, by utilizing the polar coordinate transformation based dimension reduction method, a numerical analysis method for the hybrid reliability model are provided. At last, several application cases are presented to prove the effectiveness of the proposed method for the reliability analysis under hybrid uncertainty. The comparisons between the results of the proposed method and the Monte Carlo simulation also illustrate the merit of this method.

**Keywords**: Hybrid reliability analysis; Uncertainty theory; Sub-additive measure; Mixed uncertainties

## 1. Introduce

In practice, there are various uncertainties involved in the environment and operation process which should be comprehensively considered in the reliability analysis of products [1]. Considering the existence of uncertainties at the design stage can help avoid the suboptimal design and achieve a better cost-effectiveness [2]. Generally, the uncertainty consists of two types: aleatory uncertainty and epistemic uncertainty [3, 4]. The former describes the inherent variation of a physical system, while the later is associated with the ignorance or imprecise knowledge about a system. In general, the aleatory uncertainty is depicted by the variables or distributions, while the epistemic uncertainty can be handled by the numerous non-probabilistic approaches, such as the evidence theory [5, 7], the interval analysis [8], the possibility theory [10], etc. Besides, the coexistence of them in a system will lead to a hybrid reliability analysis (HRA) problem. In order to uniformly consider the effects of both types of uncertainties, some mathematical frameworks have been proposed to solve this problem, e.g. fuzzy interval method [9], fuzzy probabilities method [19].

Due to the lack of enough information, sometimes the experts need to estimate the belief degree for some uncertain events that can be addressed by the subjective probability theory or the fuzzy theory. However, the overestimation or the underestimation about the events that effected by the epistemic uncertainty will cause a lager variance of the estimated belief degree than the actual frequency, so the precise subjective probability is hard to obtain. In addition, since the possibility measure is not self-dual (the sum of the fuzzy reliability index and the fuzzy failure index is not equal to 1), it might lead to counter-intuitive results in some application cases when introduced for describing belief degree [11]. Therefore, the epistemic uncertainty contained in an uncertain event cannot be perfectly described by the randomness or the fuzziness, the subjective probability theory or the fuzzy theory are not valid. So, to solve the existing problem, the new theory that has a more complete theoretic basis and more in line with the actual situation is in an urgent need in the uncertainty quantification area. And the reasonable measure metrics for hybrid reliability analysis of the systems under the mixed uncertainties are also needed.

In this paper, we introduce a new mathematical theory named uncertainty theory to deal with such imprecise quantities—epistemic uncertainty. This theory was proposed by Liu et al [12], which can be used to define the belief degree of the uncertain events affected by epistemic uncertainty. It has been proved that uncertainty theory is a powerful and effective method to quantify the effect of the uncertain information [11, 14]. As the theoretical basis for uncertainty theory, some concepts about the uncertain variable, the distribution, and the uncertain process were provided in [6]. Then considering the situation that aleatory and epistemic uncertainty coexist, the uncertain variable and random variable can be adopted as the measures for them respectively [20]. In this cases, the chance theory [21~23] provides a theoretical framework to consider both the uncertain variable and the random variable and defines an uncertain random variable to describe this phenomenon. Until now, the applications of the uncertainty theory and the chance theory have made achievements in various fields, including the new reliability metrics defined based on the belief degree [14, 15, 31], reliability analysis methods for the uncertain events under the epistemic uncertainty [2, 13, 30], the decision making under uncertain environment [18], uncertain finance [16, 17], and the hybrid reliability analysis problem [24~26]. The fast development and the wide application of the uncertainty theory have illustrated its potential to address the quantification problem of the aleatory and epistemic uncertainty.

Based on the uncertainty theory, this paper proposed a unified hybrid reliability analysis method for the system under mixed uncertainty, and to improve the calculation efficiency, a numerical method for this hybrid reliability model is constructed by utilizing the dimension reduction method based on the polar coordinate transformation [27] and the Hasofer-Lind-Rackwitz-Fiessler (HLRF) [28] method. For the hybrid uncertainty analysis problem, the result of the proposed method has been compared with that of other commonly used methods, which have proved the effectiveness and advantages of this method. Overall, this method can handle the quantification problem of the hybrid uncertainty reasonably and provide us some new perspectives to understand the epistemic uncertainty.

This paper is organized as follows. Section 2 provides some mathematical concepts of the uncertainty theory. Section 3 presents the unified hybrid uncertainty quantitative method. In Section 4, the procedures of the numerical method for the proposed hybrid reliability analysis model are derived based on the polar coordinate transformation. Then several cases are given in Section 5 to demonstrate the rationality and practicability of the proposed method.

## 2. Basic theoretical concepts

### 2.1 Uncertainty theory: A theory for modeling the epistemic uncertainty

In this paper, the impact of the epistemic uncertainty on the system can be quantified by the uncertainty measure $M\{\cdot\}$ that is actually a belief degree of the occurrence of an uncertain event [6]. And the independence of the uncertain variables $\tau_1, \tau_2, ..., \tau_n$ can be defined as follows: If and only if

$$M\left\{\bigcup_{i=1}^{n}(\tau_i \in B_i)\right\} = \bigvee_{i=1}^{n} M\{\tau_i \in B_i\} \tag{1}$$

for any Borel sets $B_1, B_2, ..., B_n$ of real numbers.

Then for the uncertainty distribution which can be used to quantify the incomplete knowledge represented by the uncertain variables, it can be defined as follows

$$\Upsilon(x) = M\{\tau \leq x\} \tag{2}$$

Several kinds of uncertainty distribution have been defined by Liu [6]. And for more information, please refer to the book *Uncertainty Theory*.

## 2.2 Chance theory: A mixture of probability theory and uncertainty theory

For the cases that the aleatory and epistemic uncertainty coexist in one system, the chance measure can provide a combination framework for the probability measure and the uncertainty measure which can be used as the belief degree of the occurrence of an events influenced by the hybrid uncertainty [21]. The uncertain random variable $\xi$ can be defined as a random element with an "uncertain variable" value. Then the chance measure on the chance space $(\Gamma, \mathcal{L}, M) \times (\Omega, \mathcal{A}, \Pr)$ can be defined as

$$Ch\{\xi \in B\} = \int_0^1 \Pr\{\omega \in \Omega | M\{\gamma \in \Gamma | \xi(\gamma, \omega) \in B\} \geq x\} dx \tag{3}$$

where $B$ is a Borel set of real numbers. Then the chance distribution of $\xi$ is defined by

$$\Phi(x) = Ch\{\xi \leq x\} \tag{4}$$

for any $x \in \Re$. Its actual meaning is similar to that of the uncertainty measure.

As the definitions in [29], using the independent random variables $\eta_1, \eta_2, ..., \eta_m$ and the independent uncertain variables $\tau_1, \tau_2, ..., \tau_n$, we can construct an uncertain random variable

$$\xi = f(\eta_1, \eta_2, ..., \eta_m, \tau_1, \tau_2, ..., \tau_n) \tag{5}$$

and if $\xi$ is strict increase with respect to $\tau_1, \tau_2, ..., \tau_k$ and strict decrease with respect to $\tau_{k+1}, \tau_{k+2}, ..., \tau_n$, then using the probability distributions $\Psi_1, \Psi_2, ..., \Psi_m$ for the random variables and the uncertainty distributions $\Upsilon_1, \Upsilon_2, ..., \Upsilon_n$ for the uncertain variables, its chance distribution can be defined as

$$\Phi(x) = \int_{\Re^m} F(x; \eta_1, \eta_2, ..., \eta_m) d\Psi_1(\eta_1) d\Psi_2(\eta_2) ... d\Psi_m(\eta_m) \tag{6}$$

where $F(x; \eta_1, \eta_2, ..., \eta_m)$ is the uncertainty distribution of the uncertain variable $f(\eta_1, \eta_2, ..., \eta_m, \tau_1, \tau_2, ..., \tau_n)$ for any given real numbers $\eta_1, \eta_2, ..., \eta_m$. On the one hand, when $\Upsilon_1, \Upsilon_2, ..., \Upsilon_n$ are continuous, $F(x; \eta_1, \eta_2, ..., \eta_m)$ can be determined by

$$F(x; \eta_1, \eta_2, ..., \eta_m) = \sup_{f(\eta_1, \eta_2, ..., \eta_m, \tau_1, ..., \tau_n) = x} \left( \min_{1 \leq i \leq k} \Upsilon_i(\tau_i) \wedge \min_{k+1 \leq i \leq n} (1 - \Upsilon_i(\tau_i)) \right) \tag{7}$$

On the other hand, if $\Upsilon_1, \Upsilon_2, ..., \Upsilon_n$ are regular uncertainty distributions, then $F(x; \eta_1, \eta_2, ..., \eta_m)$ is the root $\alpha$ of the equation

$$f(\eta_1, \eta_2, ..., \eta_m, \Upsilon_1^{-1}(\alpha), ..., \Upsilon_k^{-1}(\alpha), \Upsilon_{k+1}^{-1}(1-\alpha), ..., \Upsilon_n^{-1}(1-\alpha)) = x \tag{8}$$

# 3. A unified hybrid reliability analysis model

In reliability engineering, the limit-state function (LSF) can be used to describe the performance response that influenced by the uncertain parameters of the system, and numerous methods can be used to construct an LSF. In this paper, we consider the LSF consists of the uncertain variables and the random variables, which can be written as

$$G = f(\mathbf{X}, \mathbf{Y}) \tag{9}$$

where $\mathbf{X} = [\eta_1, \eta_2, ..., \eta_m]^T$ is an $m$-dimensional vector consists of the independent random variables with the probability distributions $\Psi_1, \Psi_2, ..., \Psi_m$, and $\mathbf{Y} = [\tau_1, ..., \tau_n]^T$ is an $n$-dimensional vector consists of the independent uncertain variables with the uncertainty distributions $\Upsilon_1, \Upsilon_2, ..., \Upsilon_n$. Then the performance response $G > 0$ indicates the system is safe, $G \leq 0$ indicates the system is in danger. Similar to the reliability metrics defined in [31], the belief degree of the system failure is defined as

$$F_H = Ch\{G \leq 0\} = Ch\{f(\mathbf{X},\mathbf{Y}) \leq 0\} \tag{10}$$

Then, the reliability metric can be formulated as

$$R_H = 1 - F_H = Ch\{f(\mathbf{X},\mathbf{Y}) > 0\} \tag{11}$$

If the limit state function $f(\eta_1,...,\eta_m,\tau_1,...,\tau_n)$ satisfy the conditions of the chance distribution defined in Section 2.2, then the formula of the reliability metric $R_H$ can be rewritten as

$$R_H = \int_{\Re^m} F(x;\eta_1,\eta_2,...,\eta_m) d\Psi_1(\eta_1)...d\Psi_m(\eta_m) \tag{12}$$

where $x = 0$, it means the limit state of the system between the failure and the safety. And $F(0;\eta_1,\eta_2,...,\eta_m)$ can be determined by

$$F(0;\eta_1,\eta_2,...,\eta_m) = \sup_{f(\eta_1,\eta_2,...,\eta_m,\tau_1,...,\tau_n)=0} \left( \min_{1 \leq i \leq k}(1 - \Upsilon_i(\tau_i)) \wedge \min_{k+1 \leq i \leq n} \Upsilon_i(\tau_i) \right) \tag{13}$$

or $F(0;\eta_1,\eta_2,...,\eta_m)$ is the root $\alpha$ of the equation

$$f\left(\eta_1,\eta_2,...,\eta_m, \Upsilon_1^{-1}(1-\alpha),..., \Upsilon_k^{-1}(1-\alpha), \Upsilon_{k+1}^{-1}(\alpha),..., \Upsilon_n^{-1}(\alpha)\right) = 0 \tag{14}$$

Sometimes, the equation may not have a root [6]. In this case, if

$$f\left(\eta_1,\eta_2,...,\eta_m, \Upsilon_1^{-1}(1-\alpha),..., \Upsilon_k^{-1}(1-\alpha), \Upsilon_{k+1}^{-1}(\alpha),..., \Upsilon_n^{-1}(\alpha)\right) < 0 \tag{15}$$

for all $\alpha$, then we set the root $\alpha = 0$; and if

$$f\left(\eta_1,\eta_2,...,\eta_m, \Upsilon_1^{-1}(1-\alpha),..., \Upsilon_k^{-1}(1-\alpha), \Upsilon_{k+1}^{-1}(\alpha),..., \Upsilon_n^{-1}(\alpha)\right) > 0 \tag{16}$$

for all $\alpha$, then we set the root $\alpha = 1$.

The proof of this equation is given in *Appendix A (A.1)*.

On the one hand, if the system is only affected by the aleatory uncertainty, so the limit state function G containing only random variables $\eta_1, \eta_2, ..., \eta_m$. Then the reliability metric is

$$\begin{aligned} R_H &= Ch\{f(\mathbf{X}) > 0\} \\ &= \Pr[f(\eta_1,...,\eta_m) > 0] \\ &= \int ... \int_{f(\eta_1,...,\eta_m)>0} p_\eta(\eta_1,...,\eta_m) d\eta_1...d\eta_m \\ &= R_P \end{aligned} \tag{17}$$

where $\Pr[\cdot]$ is the probability measure. So, the $R_H$ degenerates to the probabilistic reliability metric $R_P$. The proof of this equation is given in *Appendix A (A.2)*.

On the other hand, if the system is only affected by the epistemic uncertainty, and the limit state function G containing only uncertain variables $\tau_1, \tau_2, ..., \tau_n$. Then the reliability metric is

$$\begin{aligned} R_H &= Ch\{f(\mathbf{Y}) > 0\} \\ &= M[f(\tau_1,...,\tau_n) > 0] \\ &= R_M \end{aligned} \tag{18}$$

So, the $R_H$ degenerates to the uncertainty theory based reliability metric $R_M$, which is consistent with the uncertainty reliability metric proposed in [13]. The proof of this equation is given in *Appendix A (A.3)*.

In summary, the reliability metric $R_H$ of the proposed unified hybrid reliability analysis model can well depict the state of the system under hybrid uncertainty, and also applicable to the situations that only one type of uncertainty exists in the system. Therefore, this chance theory based reliability analysis model can be used as a more general method for the system assessment when the aleatory uncertainty and epistemic uncertainty coexist.

# 4. A numerical analysis method for the proposed unified reliability model

The chance theory based reliability metric has been proved to be a general index to measure the reliability of the structure under hybrid aleatory and epistemic uncertainties. However, in engineering practice, the limit state function $f(\eta_1,\eta_2,...,\eta_m,\tau_1,\tau_2,...,\tau_n)$ generally does not always meet the conditions of strict increase or decrease with respect to the uncertain variable $\tau_1,\tau_2,...,\tau_n$. To solve this problem, a novel numerical analysis method is proposed to improve the practicability and computational efficiency of the constructed hybrid reliability model, it is mainly based on the dimension reduction method of the polar coordinate transformation proposed by Hurtado [27]. The polar coordinate transformation is based on the two nonlinear features: the distance to the origin and the cosine of the angle they make with the design point unit vector $\boldsymbol{\alpha}$. The two features are given as follows:

$$v_1 = r = \|\mathbf{u}\|_2 \tag{19}$$

$$v_2 = \cos\theta = \frac{(\mathbf{u},\boldsymbol{\alpha})}{\|\mathbf{u}\|\|\boldsymbol{\alpha}\|} = \frac{(\mathbf{u},\boldsymbol{\alpha})}{v_1} \tag{20}$$

$$\boldsymbol{\alpha} = \frac{\nabla g(\mathbf{u}^*)}{\|\nabla g(\mathbf{u}^*)\|} \tag{21}$$

where $\mathbf{u}=(u_1,u_2,...,u_n)$ is the standard independent Gaussian vector, and $\mathbf{u}^*$ is the design point. In a reliability analysis problem, the design point $\mathbf{u}^*$ can be calculated by solving the optimization problem [32,33]

$$\begin{aligned} &\text{find } \mathbf{u}^* \\ &\min \ \beta = \|\mathbf{u}\| \\ &\text{s.t. } g(\mathbf{u}) = 0 \end{aligned} \tag{22}$$

where $\beta$ is the reliability metric. For the large number $n$, $v_1$ obeys a Chi density function [27]:

$$\phi_1(v_1;n) = \frac{2^{1-n/2} v_1^{(n-1)}}{\Gamma(n/2)} \exp\left(-\frac{v_1^2}{2}\right), \ v_1 > 0 \tag{23}$$

and $v_2$ obeys the following distribution:

$$\phi_2(v_2) = \frac{\sin^{n-2}(\arccos v_2) + \sin^{n-2}(\pi - \arccos v_2)}{\sqrt{1-v_2^2} \int_0^\pi \sin^{n-2}\alpha d\alpha}, \ -1 < v_2 < 1, n \geq 2. \tag{24}$$

Hurtado [27,32] has proved that the two new variables are independent, because the norm of $\mathbf{u}$ does not depend on its angle with respect to $\boldsymbol{\alpha}$ and, in its turn, the angle can be measured with respect to any other vector co-linear with $\mathbf{u}$.

## 4.1 The polar transformation of an HRA problem

As above mentioned, the performance response of a HRA problem can be described by a LSF $G = f(\mathbf{X},\mathbf{Y})$, in this section, we adopt the assumptions that $\mathbf{X} = [x_1,x_2,...,x_m]^T$ is an $m$-dimensional vector consists of the independent random variables, and $\mathbf{Y} = [y_1,y_2,...,y_n]^T$ is an $n$-dimensional vector consists of the independent uncertain variables effected by the epistemic uncertainty and we just know the lower and upper bounds of them. By utilizing the standard Gaussian transformation and the normalized interval transformation, the random variables and the uncertain variables can be transformed into the standard independent Gaussian vector $\mathbf{u} = (u_1,u_2,...,u_m) \sim N(0,1)$ and the normalized independent vector $\boldsymbol{\delta} = (\delta_1,\delta_2,...,\delta_n) \in [-1,1]$, then the LSF can be rewritten as

$$f(\mathbf{X},\mathbf{Y})=f(T_1(\mathbf{u}),T_2(\boldsymbol{\delta}))=f(\boldsymbol{\omega}) \tag{25}$$

where $T_1$ represents the standard Gaussian transformation and $T_2$ represents the normalized interval transformation. Therefore, this HRA problem is defined in the $m+n$ dimensional space $\boldsymbol{\omega}$.

Then, we transform this LSF to the problem in the polar space through the polar transformation method:

$$v_1 = \|\boldsymbol{\omega}\|_2 = \sqrt{\sum_{i=1}^{m} u_i^2 + \sum_{j=1}^{n} \delta_j^2} = \sqrt{\xi_1 + \xi_2} \tag{26}$$

$$\xi_1 = \sum_{i=1}^{m} u_i^2, \quad \xi_2 = \sum_{j=1}^{n} \delta_j^2 \in [0,n] \tag{27}$$

where $\xi_1 \sim \chi^2(m)$, and the $\xi_2 \in [0,n]$ has the 0 and $n$ as the lower and upper bounds, it can be described by the uncertainty distribution. In addition,

$$v_2 = \cos\theta = \frac{(\boldsymbol{\omega},\boldsymbol{\alpha})}{\|\boldsymbol{\omega}\|\|\boldsymbol{\alpha}\|} = \xi_3 \tag{28}$$

$$\boldsymbol{\alpha} = \frac{\nabla g(\boldsymbol{\omega}^*)}{\|\nabla g(\boldsymbol{\omega}^*)\|} \tag{29}$$

where $\boldsymbol{\omega}^*$ is the design point determined by the optimization problem

$$\begin{aligned}
\min \quad & \beta = \|\boldsymbol{\omega}\| \\
\text{s.t.} \quad & g(\boldsymbol{\omega}) = 0, \\
& \mathbf{u} \sim N(0,1), \\
& \boldsymbol{\delta} \sim \mathcal{L}(-1,1).
\end{aligned} \tag{30}$$

So the two features of the polar coordinate transformation can be transformed to three parameters $\xi_1,\xi_2,\xi_3$, and the distributions of these parameters are shown as follows:

1) $\xi_1 = \sum_{i=1}^{m} u_i^2 \sim \chi^2(m)$ is the random variable obeys a Chi-squared distribution:

$$\phi_1(\xi_1) = \frac{\xi_1^{(m/2-1)}}{2^{m/2}\Gamma(m/2)}\exp\left(-\frac{\xi_1}{2}\right), \quad v_1 > 0 \tag{31}$$

2) $\xi_2 = \sum_{j=1}^{n} \delta_j^2 \in [0,n]$ is the uncertain parameters with the known lower and upper bounds, which can be described by the uncertainty distribution.

3) For $\xi_3 = v_2 = \cos\theta$, as mentioned before, it obeys the following distribution:

$$\phi_3(\xi_3) = \frac{\sin^{n-2}(\arccos\xi_3) + \sin^{n-2}(\pi - \arccos\xi_3)}{\sqrt{1-\xi_3^2}\int_0^\pi \sin^{n-2}\alpha\, d\alpha}, \quad -1 < \xi_3 < 1, n \geq 2. \tag{32}$$

So, we can know that $\xi_1,\xi_3$ are random variables, $\xi_2$ is an uncertain variable. Then the $m+n$ dimensional HRA problem can be defined in terms of two random variables and one uncertain variable. And the LSF $f(\boldsymbol{\omega})$ of the HRA problem is transformed from the space $\boldsymbol{\omega}$ to the polar space and can be represented as

$$G = g(\xi_1,\xi_2,\xi_3) \tag{33}$$

## 4.2 The proposed numerical analysis method

In this section, to calculate the reliability of the system by the LSF $G = g(\xi_1,\xi_2,\xi_3)$ in the polar space, we first need to process the limit state function by the first-order Taylor expansion, then we can construct the first-order reliability method [34,35] to get the reliability. The first-order Taylor expansion of the LSF at a design point $\boldsymbol{\omega}^*$ is

$$f(\boldsymbol{\omega}) \approx f(\boldsymbol{\omega}^*) + \sum_{i=1}^{m} \left.\frac{\partial f}{\partial u_i}\right|_{\boldsymbol{\omega}^*} (u_i - u_i^*) + \sum_{j=1}^{n} \left.\frac{\partial f}{\partial \delta_j}\right|_{\boldsymbol{\omega}^*} (\delta_j - \delta_j^*) \tag{34}$$

where $\boldsymbol{\omega}^* = (u_1^*, u_2^*, ..., u_m^*, \delta_1^*, \delta_2^*, ..., \delta_n^*)$ can be obtained by the optimization problem in Section 4.1. Then, according to the polar transformation, let

$$d = \frac{f(\boldsymbol{\omega}^*) - \sum_{i=1}^{m}(\partial f/\partial u_i)\big|_{\boldsymbol{\omega}^*} u_i^* - \sum_{j=1}^{n}(\partial f/\partial \delta_j)\big|_{\boldsymbol{\omega}^*} \delta_j^*}{\left[\sum_{i=1}^{m}\left((\partial f/\partial u_i)\big|_{\boldsymbol{\omega}^*}\right)^2 + \sum_{j=1}^{n}\left((\partial f/\partial \delta_j)\big|_{\boldsymbol{\omega}^*}\right)^2\right]^{1/2}} \tag{35}$$

$$v_1 = \|\boldsymbol{\omega}\|_2 = \sqrt{\sum_{i=1}^{m} u_i^2 + \sum_{j=1}^{n} \delta_j^2} = \sqrt{\xi_1 + \xi_2} \tag{36}$$

$$v_2 = \cos\angle(\boldsymbol{\omega}^*, \boldsymbol{\alpha}) = \frac{\sum_{i=1}^{m}(\partial f/\partial u_i)\big|_{\boldsymbol{\omega}^*} u_i + \sum_{j=1}^{n}(\partial f/\partial \delta_j)\big|_{\boldsymbol{\omega}^*} \delta_j}{v_1\left[\sum_{i=1}^{m}\left((\partial f/\partial u_i)\big|_{\boldsymbol{\omega}^*}\right)^2 + \sum_{j=1}^{n}\left((\partial f/\partial \delta_j)\big|_{\boldsymbol{\omega}^*}\right)^2\right]^{1/2}} = \xi_3 \tag{37}$$

where

$$\boldsymbol{\alpha} = \frac{\boldsymbol{\omega}^*}{\|\boldsymbol{\omega}^*\|} = \frac{\left((\partial f/\partial \omega_1)\big|_{\boldsymbol{\omega}^*}, ..., (\partial f/\partial \omega_{m+n})\big|_{\boldsymbol{\omega}^*}\right)}{\left[\sum_{k=1}^{m+n}\left((\partial f/\partial \omega_k)\big|_{\boldsymbol{\omega}^*}\right)^2\right]^{1/2}} \tag{38}$$

Then the Equation (34) can be derived as follows

$$\hat{f}(\boldsymbol{\omega}) \approx f(\boldsymbol{\omega}^*) + \sum_{i=1}^{m} \left.\frac{\partial f}{\partial u_i}\right|_{\boldsymbol{\omega}^*} (u_i - u_i^*) + \sum_{j=1}^{n} \left.\frac{\partial f}{\partial \delta_j}\right|_{\boldsymbol{\omega}^*} (\delta_j - \delta_j^*) = d*D + D*\sqrt{\xi_1 + \xi_2}*\xi_3 = g(\xi_1, \xi_2, \xi_3) \tag{39}$$

where $D = \left[\sum_{i=1}^{m}\left((\partial f/\partial u_i)\big|_{\boldsymbol{\omega}^*}\right)^2 + \sum_{j=1}^{n}\left((\partial f/\partial \delta_j)\big|_{\boldsymbol{\omega}^*}\right)^2\right]^{1/2}$, and $D > 0$.

Therefore, based on the polar transformation and the Taylor expansion, the $m+n$ dimensional LSF of the HRA problem can be derived as a three dimension problem in the polar space, shown as follows

$$G = g(\xi_1, \xi_2, \xi_3) = d*D + D*\sqrt{\xi_1 + \xi_2}*\xi_3 \tag{40}$$

where the safe domain is $\Omega: d + \sqrt{\xi_1 + \xi_2}*\xi_3 > 0$, and the failure domain is $\bar{\Omega}: d + \sqrt{\xi_1 + \xi_2}*\xi_3 \leq 0$.

Furthermore, according to the unified hybrid reliability analysis model proposed in Section 3, the reliability of the HRA problem can be formulated as

$$R_H = Ch\{g(\xi_1, \xi_2, \xi_3) > 0\} \tag{41}$$

where $\xi_1, \xi_3$ are random variables with probability distributions $\Psi_1, \Psi_3$, and $\xi_2$ is an uncertain variable with uncertainty distribution $\Upsilon$. In addition, as the LSF $g(\xi_1, \xi_2, \xi_3)$ contains one uncertain variable and two random variables, according to the Equation (12), then the reliability $R_H$ can be calculated as

$$R_H = \int_{\Re^2} F(0; \xi_1, \xi_3) d\Psi_1(\xi_1) d\Psi_3(\xi_3) \tag{42}$$

## 4.3 Algorithm for the proposed numerical analysis method

To improve the efficiency and applicability of the numerical method proposed in Section 4.2, by utilizing an uncertainty theory based HLRF method (uHLRF) for searching the design point and the sequential optimization strategy, the HRA problem under mixed uncertainties can be solved with the algorithm as follows:

**Step 1**: For the random variables $x_1, x_2, ..., x_m$ and the uncertain variables $y_1, y_2, ..., y_n$ only knows the lower and upper bounds, by utilizing the standard Gaussian transformation and the normalized interval transformation, convert $\mathbf{X} = [x_1, x_2, ..., x_m]^T$ and $\mathbf{Y} = [y_1, y_2, ..., y_n]^T$ into the standard independent Gaussian vector $\mathbf{u} = (u_1, u_2, ..., u_m) \sim N(0,1)$ and the normalized independent vector $\boldsymbol{\delta} = (\delta_1, \delta_2, ..., \delta_n) \in [-1,1]$, which respectively described by the probability distribution and the uncertainty distribution. Due to the simultaneous existence of the random variables $\mathbf{u} = (u_1, u_2, ..., u_m)$ and the uncertain variables $\boldsymbol{\delta} = (\delta_1, \delta_2, ..., \delta_n)$, we proposed an uncertainty theory based HLRF procedure to solve the optimization problem in Section 4.1. This method is a single-loop procedure, which consists of the probability analysis (PA) loop and the uncertainty analysis (UA) loop, UA is embedded in the iterative design point $\mathbf{u}_k^*$ search process.

**Step 2**: UA Loop. With the initial point $\mathbf{u}, \boldsymbol{\delta}$, the uncertainty analysis loop is conducted with the fixed $\mathbf{u}$. Once $\mathbf{u}_{k-1}^*$ is found, UA is conducted to find the design point $\boldsymbol{\delta}_k^*$ by the following model:

$$\begin{cases} \text{find} \quad \boldsymbol{\delta}_k^* \\ \min_{\boldsymbol{\delta}} \beta_{k-1} = \left\| (\mathbf{u}_{k-1}^*, \boldsymbol{\delta}) \right\| \\ \text{st.} \quad f(\mathbf{u}_{k-1}^*, \boldsymbol{\delta}) = 0 \\ \boldsymbol{\delta} \in (\Gamma, \mathcal{L}, M) \end{cases} \quad (43)$$

**Step 3**: PA Loop. With the obtained point $\boldsymbol{\delta}_k^*$, by utilizing the HLRF algorithm [33,36] to calculate the design point $\mathbf{u}_k^*$

$$\begin{cases} \beta_k = \beta_{k-1} + \dfrac{\mathbf{u}_{k-1}^*}{\left\| \nabla f(\mathbf{u}_{k-1}^*, \boldsymbol{\delta}_k^*) \right\|} \\ \mathbf{u}_k^* = -\beta_k \dfrac{\nabla f(\mathbf{u}_{k-1}^*, \boldsymbol{\delta}_k^*)}{\left\| \nabla f(\mathbf{u}_{k-1}^*, \boldsymbol{\delta}_k^*) \right\|} \end{cases} \quad (44)$$

**Step 4**: Check for convergence. If $\left\| \boldsymbol{\omega}_k^* - \boldsymbol{\omega}_{k-1}^* \right\| \leq \varepsilon$ ($\varepsilon$ is small positive number $\geq 0.0$), then $\boldsymbol{\omega}^* = \boldsymbol{\omega}_k^*$ and go to Step 5; otherwise, k=k+1, go to Step 2.

**Step 5**: Let the design point $\boldsymbol{\omega}^* = (u_1^*, ..., u_m^*, \delta_1^*, ..., \delta_n^*)$, and calculate the $(\partial f / \partial \omega_k)\big|_{\boldsymbol{\omega}^*}$ of the limit state function $f(\boldsymbol{\omega})$ at the point $\boldsymbol{\omega}^*$.

**Step 6**: Utilizing the method proposed in section 4.2, transform the function $f(\boldsymbol{\omega})$ into $g(\xi_1, \xi_2, \xi_3) = d * D + D * \sqrt{\xi_1 + \xi_2} * \xi_3$ in the polar space, and calculate the parameters $d, D$.

**Step 7**: Then, according to Equation (12), the reliability $R_H$ can be calculated by the double integral as follows

$$R_H = \int_{\Re^2} F(0; \xi_1, \xi_3) \phi_1(\xi_1) \phi_3(\xi_3) d\xi_1 d\xi_3 \quad (45)$$

where $\Re^2$ stands for the safe domain $\Omega: d + \sqrt{\xi_1 + \xi_2} * \xi_3 > 0$.

**Step 8**: As there are only one uncertain variable contained in the limit-state function $g(\xi_1, \xi_2, \xi_3) = 0$, in order to achieve an efficient calculation, the uniformity sequential sampling strategy is conducted to handle this problem. The uncertain variable $\xi_2$ exists in the integral domain is regarded as an error term $\sigma = \sum_{j=1}^{n} \delta_j^2 \in [0, n]$ obeys the uncertainty distribution. Then $v_1 = \sqrt{\xi_1 + \xi_2} = \sqrt{\sum_{i=1}^{m} u_i^2 + \sigma}$, which can be regarded as a random variable with the disturbance term, and $v_1$ obeys the Chi distribution written as follows:

$$\phi_1(v_1) = \frac{2^{1-m/2}(v_1^2-\sigma)^{(m-1)/2}}{\Gamma(m/2)} \exp\left(-\frac{v_1^2-\sigma}{2}\right) \tag{46}$$

Then the reliability $R_H$ can be calculated as follows

$$R_H = \int_{\Re^2} F(0; v_1^2-\sigma, \xi_3)\phi_1(v_1)\phi_3(\xi_3)dv_1d\xi_3 \tag{47}$$

where $\Re^2$ stands for the safe domain $\Omega: d+v_1*\xi_3>0, v_1>\sqrt{\sigma}$. Therefore, the formula of the reliability $R_H$ is transformed to the double integral of $v_1, \xi_3$. The error term $\sigma$ is processed by the sequential sampling strategy under the uncertainty distribution, and the reliability can be calculated by the Equation (47). Then, according to the lower and upper bounds of the uncertain variable $\sigma$, the range of the reliability $R_H$ can be obtained.

The flowchart of this algorithm for the hybrid reliability analysis is depicted in Figure 1.

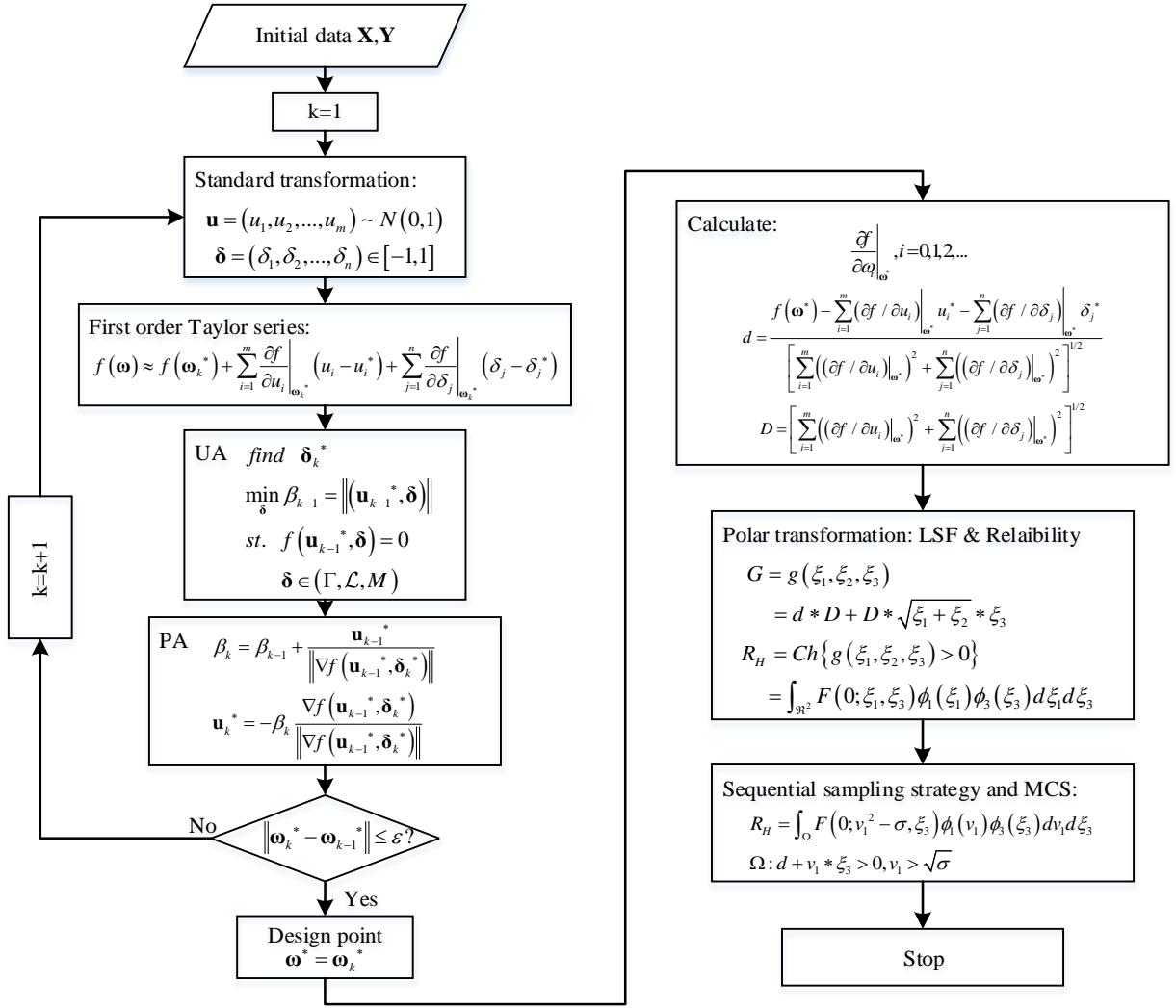

Figure 1. Flowchart of the proposed numerical analysis method

# 5. Numerical examples and discussions

## 5.1 Numerical Example

The following linear limit-state function is considered:

$$g(u,\delta) = 1 - f(u,\delta) = 1 - (\sum_{i=1}^{m} u_i + \sum_{j=1}^{n} \delta_j)/(m+n)$$

where the independent random variables $u_i (i=1,2,...,m)$ obeying the standard normal distribution $N(0,1)$ and the normalized independent vector $\delta_j (j=1,2,...,n)$ have the range of $[-1,1]$. By utilizing the method that proposed in Section 4.3, the calculated reliability results with different $m$ and $n$ are shown in the following table. Besides, for comparison purposes, a Monte Carlo simulation involving the generation of 1000000 random realizations of input variables **u** and the computation of the reliability for each of them with 1000000 samples of variables **δ** with the uniform distributions was performed.

From the results shown in Table 1 and Fig.2, the trend of the failure rate $F_H$ obtained from the proposed method is consistent with that calculated by the MCS method. With the increase of the number of the random variables, the epistemic uncertainty decreases in the function $f(u,\delta)$, the failure rate of the limit state function $g(u,\delta)$ increases. Furthermore, compared with the MCS result, we can see that the proposed method achieves good accuracy, and the failure rate $F_H$ is larger than that obtained from MCS method. Because in this MCS, the uncertain variables are treated as the uniform variables, the epistemic uncertainties are ignored which have negative effect on the accuracy of the results, and lead to an overestimate of the reliability level. Therefore, when the epistemic uncertainty and aleatory uncertainty simultaneously coexist in a structure system, ignoring the epistemic uncertainty will cause an adverse effect on the results, so the method based on the probability theory is not suitable to handle this HRA problem. On the contrary, the $R_H$ results obtained from the proposed method can provide a more realistic references for the hybrid reliability analysis problem.

**Table 1** Analysis results of the proposed method and the MCS

| m | n | Failure rate | |
|---|---|---|---|
| | | $F_H$ | MCS (confidence 95%) |
| 1 | 9 | [5.736e-7, 8.993e-6] | [5.750e-8, 1.425e-7] |
| 3 | 7 | [2.132e-5, 1.039e-4] | [1.050e-6, 1.045e-5] |
| 5 | 5 | [1.441e-4, 3.174e-4] | [3.256e-5, 6.294e-5] |
| 7 | 3 | [6.863e-4, 9.075e-4] | [1.701e-4, 2.233e-4] |
| 9 | 1 | [4.015e-3, 4.109e-3] | [5.135e-4, 5.653e-4] |

Note: m and n denote the number of random variables and the uncertain variables respectively.

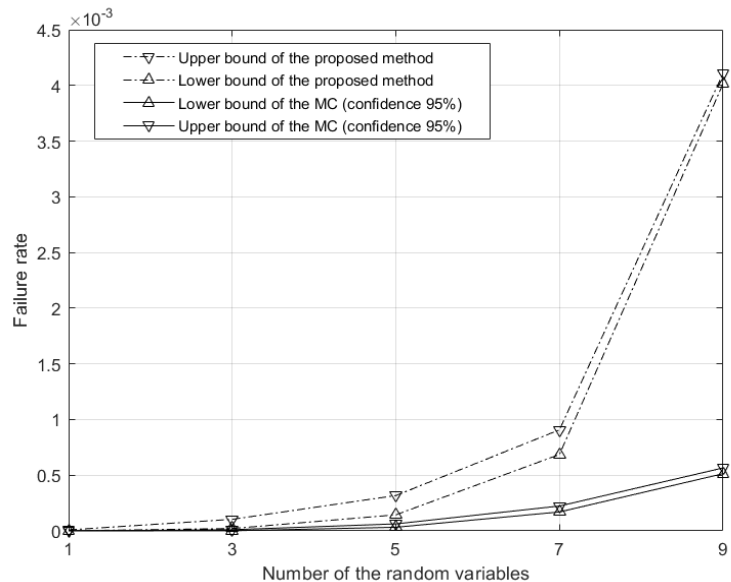

Figure 2. Comparison of the results

## 5.2 A Crank-slider Mechanism

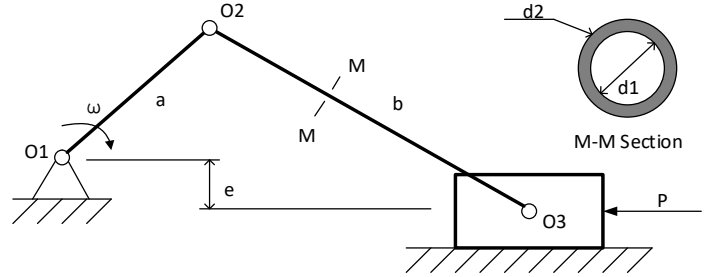

Figure 3. A crank-slider mechanism

A crank-slider mechanism as shown in Fig.3 is investigated, which is modified from the case in [5]. The limit-state function is defined by the difference between the material strength and the maximum stress. The function is given by

$$g(a,b,P,e) = S_m - S$$

and

$$S = \frac{4P(b-a)}{\pi\left(\sqrt{(b-a)^2 - e^2} - \mu e\right)\left(d_2^2 - d_1^2\right)}$$

where $\mu$ is the coefficient of friction between the ground and the slider, and $\mu$ is time-dependent $\mu = 0.30 + 0.002*t$. Due to the manufacturing errors and the harsh environment, the precise distributions of $a, b, P, e$ are unknown, only their upper and lower bounds are available from the suggestions from experts. The meanings and distributions of the variables are given in Table 2. The results of the MCS and the proposed method are shown in Figure 4. When the time is t=0, the failure rate obtained by the MCS method is 0.06873, and the result calculated by the proposed method is [0.05152, 0.07276]; t=40, the failure rate results obtained from the MCS and the proposed method are 0.18423 and [0.21260, 0.25600]. The comparisons show that the results obtained by the proposed method is more conservative. From the result, we can know that the trends of the results of the MCS method and the proposed method are consistent. As the time $t$ increases, the coefficient of friction $\mu$ becomes larger, the reliability level of the structure decreases. This can be explained by the fact that the larger friction coefficient will lead to faster wear of the structure, which will introduce more influence of the epistemic uncertainty into the reliability model. So, with the time $t$ increases, the increasing speed of the failure rate becomes larger. This case illustrates that the proposed method can depict the effect of the epistemic uncertainty in a more realistic way and avoid the overestimation of the system reliability.

**Table 2** Variables of the crank-slider mechanism.

| Variables | Meaning | Parameter 1 | Parameter 2 | Style |
|---|---|---|---|---|
| $d_1$ (mm) | inner diameter of the coupler | 10 | 0.5 | Normal |
| $d_2$ (mm) | outer diameter of the coupler | 20 | 0.8 | Normal |
| $S_m$ (MPa) | the yield strength of the coupler | 1.98 | 0.1 | Normal |
| $a$ (mm) | length of the crank | 94 | 106 | Uncertain variable |
| $b$ (mm) | length of the coupler | 295 | 305 | Uncertain variable |
| $P$ (kN) | the external force | 240 | 260 | Uncertain variable |
| $e$ (mm) | the offset | 122 | 128 | Uncertain variable |

Note: For the probability variable, parameters 1 and 2 denote the mean and standard deviation respectively. For the uncertain variable, parameters 1 and 2 denote the lower and upper bound respectively.

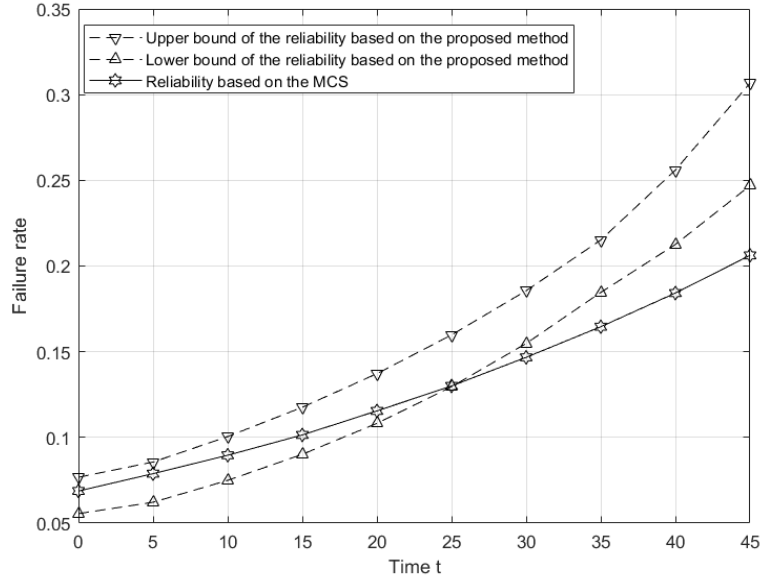

Figure 4. Comparison of the results

## 5.3 A Cantilever Tube

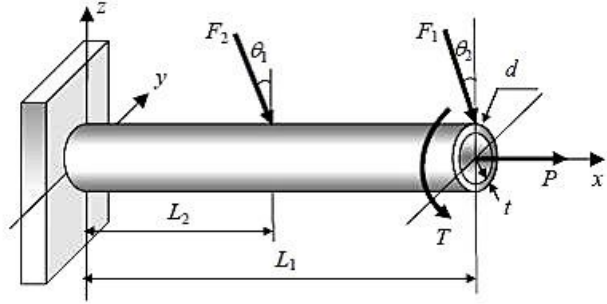

Figure 5. The cantilever tube

The cantilever tube is subjected to external forces $F_1, F_2, P$ and torsion $T$. The LSF proposed in [37] is expressed as follows:

$$g = S_y - \sigma_{max} + \omega$$

where $S_y$ is the yield strength, $\omega$ is the added Gaussian noise to simulate a more complex LSF, which can be used to assess the robustness and practicability of the method, and the $\sigma_{max}$ is the maximum von Mises stress on the top surface of the tube at the origin, which is given by

$$\sigma_{max} = \sqrt{\sigma_x^2 + 3\tau_{xy}^2}$$

the stress $\sigma_x$ is determined by

$$\sigma_x = \frac{P + F_1 \sin\theta_1 + F_2 \sin\theta_2}{A} + \frac{Mc}{I}$$

where $A$ is the area of the tube, $M$ is the bending moment, which are calculated by

$$A = (\pi/4)\left[d^2 - (d - 2t)^2\right]$$

$$M = (F_1 L_1 \cos\theta_1 + F_2 L_2 \cos\theta_2)(d/2)$$

and the torsional stress $\tau_{xy}$ is given by

$$\tau_{xy} = \frac{Td}{4I}$$

where $I = (\pi/64)\left[d^2 - (d-2t)^4\right]$.

The random variables and uncertain variables are presented in Table 3 and Table 4. By utilizing the algorithm proposed in Section 4.3, the failure rate $F_H$ of the structure is [2.859e-3, 5.790e-3], and the result with 95% confidence calculated by the MCS method is [3.8253e-4, 6.8707e-4]. From the results, we can know that the proposed method achieves a narrower range of the reliability evaluation than the MCS, and this indicates that the proposed method can better handle the epistemic uncertainty. Since the existence of the epistemic uncertainty has a nonnegligible effect on the accuracy of the hybrid reliability analysis results. The proposed method can deal with the influence of the epistemic uncertainty in a more realistic way, and provide a more suitable assessment to the reliability level of the structure. This case proves that for the high dimensional nonlinear problems under the mixed uncertainties, the proposed method is also applicable.

**Table 3** Random variables of the cantilever tube problem.

| Variables | Parameter 1 | Parameter 2 | Style |
|---|---|---|---|
| r (mm) | 5 | 0.1 | Normal |
| d (mm) | 42 | 0.5 | Normal |
| L1 (mm) | 120 | 1.2 | Normal |
| L2 (mm) | 60 | 0.6 | Normal |
| Sy (mm) | 185.0 | 22.0 | Normal |
| W (mm) | 0 | 0.03 | Normal |

Note: For the probability variable, parameters 1 and 2 denote the mean and standard deviation respectively.

**Table 4** Uncertain variables of the cantilever tube problem.

| Variables | Intervals | Style |
|---|---|---|
| θ1 | [0°, 10°] | Uncertain variable |
| θ2 | [5°, 15°] | Uncertain variable |
| F1 (kN) | [12.7, 13.3] | Uncertain variable |
| F2 (kN) | [12.7, 13.3] | Uncertain variable |
| P (kN) | [21, 23] | Uncertain variable |
| T (Nm) | [85, 95] | Uncertain variable |

**Table 5** Reliability analysis results

| Method | Failure | |
|---|---|---|
| | Lower bound | Upper bound |
| The proposed method | 2.859e-3 | 5.790e-3 |
| MCS (confidence 95%) | 3.8253e-4 | 6.8707e-4 |

## 5.4 Application to a practical structure

The present method is also applied to a practical automobile frame structure, the automotive drive axle housing, which is one of the main load-bearing components of automobiles. A finite element model of the automotive drive axle housing structure is considered, as shown in Fig.6. This model contains 6890 nodes and 27925 elements. The Poisson's ratio $\mu$ of the frame and the loads $F_1, F_2$ are treated as uncertain variables with the lower and upper bounds, the density $\rho$ is treated as random variables, and their distributions are given in Table 6.

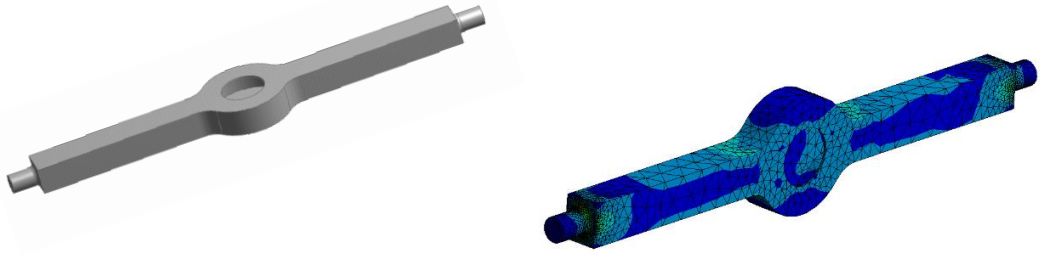

Figure 6. Automotive drive axle housing structure

According to the failure criteria of strength and stress, we can establish the limit-state function

$$g = \sigma_{max} - \sigma(F_1, F_2, \mu, \rho)$$

where $\sigma_{max}$ represents the allowable material yield stress; $\sigma(F_1, F_2, \mu, \rho)$ represents the real maximum stress, which is obtained by the finite element method (FEM). The proposed algorithm in this paper is adopted to compute the hybrid reliability. The failure probability calculated by the FEM is 0.0012 ($F_1, F_2, \mu$ are assumed to obey the uniform distributions), and the failure rate $F_H$ obtained from the propose method is [0.0697, 0.0426] ($F_1, F_2, \mu$ are treated as the uncertain variables, and assumed to obey the linear uncertainty distributions). The results of failure rate obtained by the FEM and the proposed method in this case are consistent with the results of the previous cases, which indicated that the reliability level of the structure is affected by the epistemic uncertainty, and ignoring the epistemic uncertainty can cause negative effects on the results of the HRA problem. In the FEM, the epistemic uncertainties are treated as random variables, so it leads to the overestimation of structural reliability. Obviously, in engineering practice, when the parameters cannot be precisely determined, the methods based on the probability theory are not suitable to deal with this situation. The method based on the uncertainty theory can satisfy the subadditivity characteristic of the epistemic uncertainty in theoretical, which is more suitable to handle the epistemic uncertainty. Therefore, the proposed method is applicable to deal with the HRA problems in the engineering practice.

**Table 6** Parameters of the structure

| Variables | Parameter 1 | Parameter 2 | Style |
|---|---|---|---|
| F1 (N) | 14625 | 15375 | Uncertain variable |
| F2 (N) | 14625 | 15375 | Uncertain variable |
| u | 0.2925 | 0.3075 | Uncertain variable |
| $\rho$ (kg/m^3) | 7850 | 50 | Random variable |

Note: For the random variable, parameters 1 and 2 denote the mean and standard deviation respectively. For the uncertain variable, parameters 1 and 2 denote the lower and upper bound respectively.

# 6. Conclusions

In this paper, the uncertainty theory based unified hybrid reliability analysis model and an efficient numerical method are proposed. And the proposed method achieves promising performance in the reliability analysis of the structures under hybrid uncertainty. The main contributions of this paper include two aspects:

1) A chance theory based uncertainty quantification framework is constructed to uniformly deal with the HRA problem of structures under the aleatory uncertainty and the epistemic uncertainty;

2) Proposed an efficient numerical analysis method for the proposed hybrid reliability model by using the polar coordinate transformation based dimension reduction method.

The proposed method can provide more realistic and efficient references for the hybrid reliability analysis of the structures under mixed uncertainties than the method that process the uncertain parameters by the random variables. In the future work, we will further explore the application of the uncertainty theory in time-varying reliability modeling and the reliability optimization.

# Appendix A. Proofs of Theorems

**A.1** It follows from the definition of $R_H$ and the definition of the chance distribution that

$$\begin{aligned}
R_H &= Ch\{f(\eta_1,...,\eta_m,\tau_1,...,\tau_n) > 0\} \\
&= \int_0^1 \Pr\{\omega \in \Omega | M\{f(\eta_1(\omega),...,\eta_m(\omega),\tau_1,...,\tau_n) > 0\} \geq x\} dx \\
&= \int_{\Re^m} M\{f(y_1,...,y_m,\tau_1,...,\tau_n) > 0\} d\Psi_1(y_1)...d\Psi_m(y_m) \\
&= \int_{\Re^m} G(y_1,...,y_m) d\Psi_1(y_1)...d\Psi_m(y_m)
\end{aligned}$$

where $G(y_1,...,y_m) = M\{f(y_1,...,y_m,\tau_1,...,\tau_n) > 0\}$ is the uncertainty distribution of the uncertain variable $f(y_1,y_2,...,y_m,\tau_1,\tau_2,...,\tau_n)$ for any given real numbers $y_1, y_2,..., y_m$. According to the definition of the uncertainty distribution in [6], when $\Upsilon_1, \Upsilon_2,..., \Upsilon_n$ are continuous uncertainty distributions, $G(y_1,...,y_m)$ can be determined by the following formula:

$$G(y_1,...,y_m) = \sup_{f(y_1,...,y_m,x_1,...,x_n)=0} \left( \min_{1 \leq i \leq k}(1 - \Upsilon_i(x_i)) \wedge \min_{k+1 \leq i \leq n} \Upsilon_i(x_i) \right)$$

Otherwise, when $\Upsilon_1, \Upsilon_2,..., \Upsilon_n$ are regular uncertainty distributions, $G(y_1,...,y_m)$ is the root $\alpha$ of the equation

$$f\left(y_1,...,y_m, \Upsilon_1^{-1}(1-\alpha),..., \Upsilon_k^{-1}(1-\alpha), \Upsilon_{k+1}^{-1}(\alpha),..., \Upsilon_n^{-1}(\alpha)\right) = 0$$

The theorem is thus verified.

**A.2** It follows from the definition of the uncertain random variable and the definition of $R_H$ that

$$\begin{aligned}
R_H &= Ch\left[f(\eta_1,...,\eta_m) > 0\right] \\
&= \Pr\left[f(\eta_1,...,\eta_m) > 0\right] \\
&= \int...\int_{f(\eta_1,...,\eta_m)>0} d\Psi_1(y_1)...d\Psi_m(y_m) \\
&= \int...\int_{f(\eta_1,...,\eta_m)>0} p_\eta(\eta_1,...,\eta_m) d\eta_1...d\eta_m \\
&= R_P
\end{aligned}$$

The result is thus verified.

**A.3** It follows from the definition of the uncertain random variable that

$$\begin{aligned}
R_H &= Ch\left[f(\tau_1,...,\tau_n) > 0\right] \\
&= M\left[f(\tau_1,...,\tau_n) > 0\right] \\
&= 1 - \Phi(0) \\
&= \alpha \\
&= R_M
\end{aligned}$$

then

$$\Phi(0) = 1 - \alpha$$

If $\Upsilon_1, \Upsilon_2,..., \Upsilon_n$ are regular uncertainty distributions, according to the definition of the inverse uncertainty distribution [6], then

$$\Phi^{-1}(1-\alpha) = f\left(\Upsilon_1^{-1}(1-\alpha),..., \Upsilon_k^{-1}(1-\alpha), \Upsilon_{k+1}^{-1}(\alpha),..., \Upsilon_n^{-1}(\alpha)\right) = 0$$

thus $\alpha$ is the root of the equation.

If $\Upsilon_1, \Upsilon_2,..., \Upsilon_n$ are continuous uncertainty distributions, according to the definition of the uncertainty distribution in the article [6],

$$\begin{aligned} R_H &= M\left[ f(\tau_1,...,\tau_n) > 0 \right] \\ &= 1 - \Phi(0) \\ &= \sup_{f(x_1,...,x_n)=0} \left( \min_{1 \leq i \leq k}(1 - \Upsilon_i(x_i)) \wedge \min_{k+1 \leq i \leq n} \Upsilon_i(x_i) \right) \\ &= R_M \end{aligned}$$

The result is verified.

# Acknowledgements

The authors declare that there is no conflict of interest regarding the publication of this paper.